\documentclass[aps, prl, showpacs,reprint, superscriptaddress]{revtex4-1}
\usepackage{graphicx}
\usepackage{dcolumn}
\usepackage{bm}
\usepackage{float}
\usepackage{subfigure}
\usepackage{amsmath}
\usepackage{gensymb}
\usepackage{geometry}
\usepackage{epsfig}
\usepackage{epstopdf}
\usepackage{amssymb}
\emergencystretch=\maxdimen
\begin{document}

\title{Breaking Pauli blockade via ultrafast cooling of hot electrons \\in optically-pumped graphene}

\author{Yingying Zhu}
\affiliation{National Laboratory of Solid State Microstructures, School of Physics, and Collaborative\\Innovation Center of Advanced Microstructures, Nanjing University, Nanjing 210093, China}

\author{Lianzi Liu}
\affiliation{National Laboratory of Solid State Microstructures, School of Physics, and Collaborative\\Innovation Center of Advanced Microstructures, Nanjing University, Nanjing 210093, China}

\author{Jianan Wang}
\affiliation{National Laboratory of Solid State Microstructures, School of Physics, and Collaborative\\Innovation Center of Advanced Microstructures, Nanjing University, Nanjing 210093, China}

\author{Ruwen Peng}
\thanks{Corresponding author: rwpeng@nju.edu.cn}
\affiliation{National Laboratory of Solid State Microstructures, School of Physics, and Collaborative\\Innovation Center of Advanced Microstructures, Nanjing University, Nanjing 210093, China}

\author{Dongxiang Qi}
\thanks{Corresponding author: dxqi@nju.edu.cn}
\affiliation{National Laboratory of Solid State Microstructures, School of Physics, and Collaborative\\Innovation Center of Advanced Microstructures, Nanjing University, Nanjing 210093, China}

\author{Wenzhong Bao}
\affiliation{School of Microelectronics, Fudan University, 220 Handan Road,
Shanghai, 200433, China}

\author{Renhao Fan}
\affiliation{National Laboratory of Solid State Microstructures, School of Physics, and Collaborative\\Innovation Center of Advanced Microstructures, Nanjing University, Nanjing 210093, China}

\author{Mu Wang}
\thanks{Corresponding author: muwang@nju.edu.cn}
\affiliation{National Laboratory of Solid State Microstructures, School of Physics, and Collaborative\\Innovation Center of Advanced Microstructures, Nanjing University, Nanjing 210093, China}

\date{\today\ for Version 2 \quad  }

\begin{abstract}
Pauli blockade occurs when the excited electrons fill up the states near the conduction bands and block subsequent absorption in semiconductors, and has been widely applied in mode-locking for passively-pulsed-laser systems. In this letter, we report the first direct observation that the Pauli blockade is broken by ultrafast cooling of hot electrons in optically-pumped graphene. With femtosecond spectroscopy, we demonstrate that the time scale to excite an electron ($\sim$100 fs) is of the same order as that of the electron decay via electron-electron scattering, which allows the electron excitation interplays strongly with the cooling of hot electrons. Consequently, Pauli blockade is dismissed, leading to an unconventionally enhanced optical absorption. We suggest that this effect is a universal feature of materials with simple energy level scheme, which sheds the light of ultrafast carrier dynamics in nonlinear physics and inspires the designing of new-generation of ultrafast optoelectronic devices.
\end{abstract}


\maketitle
The Pauli exclusion principle (PEP) is one of the most fundamental features in quantum world, based on which two or more identical fermions cannot simultaneously occupy the same state \cite{1,2,3}. This concept was originally proposed to elaborate electron shell structure of atoms in early 1920s, and subsequently generated enormous far-reaching impact not only to atomic physics, but to high-energy physics, condensed matter physics and beyond \cite{1,4,5,6,7}. The PEP leads to saturable absorption (SA), \emph{i.e.}, the quenching of optical absorption under high-intensity illumination \cite{8}, which is known as a universal nonlinear optical phenomenon. Energy states in a semiconductor can be occupied by large amount of induced carriers, which forbid the subsequent optical transition to these filled states due to the Pauli blockade \cite{9,10}. Saturable absorbers are practically used for stable pulse generation of solid-state lasers in both mode-locking and Q-switching regimes \cite{11}. Traditionally semiconductor saturable absorber mirrors (SESAMs) are applied to generate ultrafast pulse from continuous-wave lasers \cite{12,13,14}.  It can be characterized by a simplest two-level model in a steady state with an assumption that the longest recovery time of the system is much shorter than the pulse duration \cite{15,16}.
\setlength{\parskip}{0.15\baselineskip}

Recently a new type of saturable absorbers based on two-dimensional materials has been developed, forming a desirable platform for laser operation because of its inherently faster recovery time and compatible size to optical fibers \cite{17,18,19,20,21,22}. Both theoretical and experimental studies have been carried out on the saturation behaviour of the optically induced carrier occupation in graphene. An unconventional double-bended saturation of carrier occupation has also been presented, which is ascribed to many-particle interaction \cite{23,24}. However, so far there is no direct evidence on the ultrafast carrier dynamic processes, and the associated dynamic mechanism remains not well addressed. Identifying the ultrafast carrier dynamics will certainly inspire the designing and fabrication of high-speed functional nonlinear nano-devices.

In this letter, we report the first direct observation of breaking Pauli blockade induced by ultrafast cooling of hot electrons in an optically-pumped graphene sample. By shining a femtosecond laser beam on the sample, electrons are pumped to the empty conduction bands with high energies. These hot electrons interact each other, cool down and then, evacuate the conduction bands. Our femtosecond optical pump-probe spectroscopy demonstrates that the time scale for electron excitation ($\sim$100 fs) is of the same order as that of electron cooling via electron-electron scattering. In this way, the conduction bands are repeatedly populated-depopulated. Eventually Pauli blockade is dismissed, leading to an unconventionally enhanced optical absorption. This ultrafast dynamic process is proposed for the first time, and is experimentally verified in graphene system. We suggest that similar ultrafast dynamic process may occur in other materials with simple energy level scheme as well.

\begin{figure}[hbt]
\centering
\includegraphics[width=0.48\textwidth]{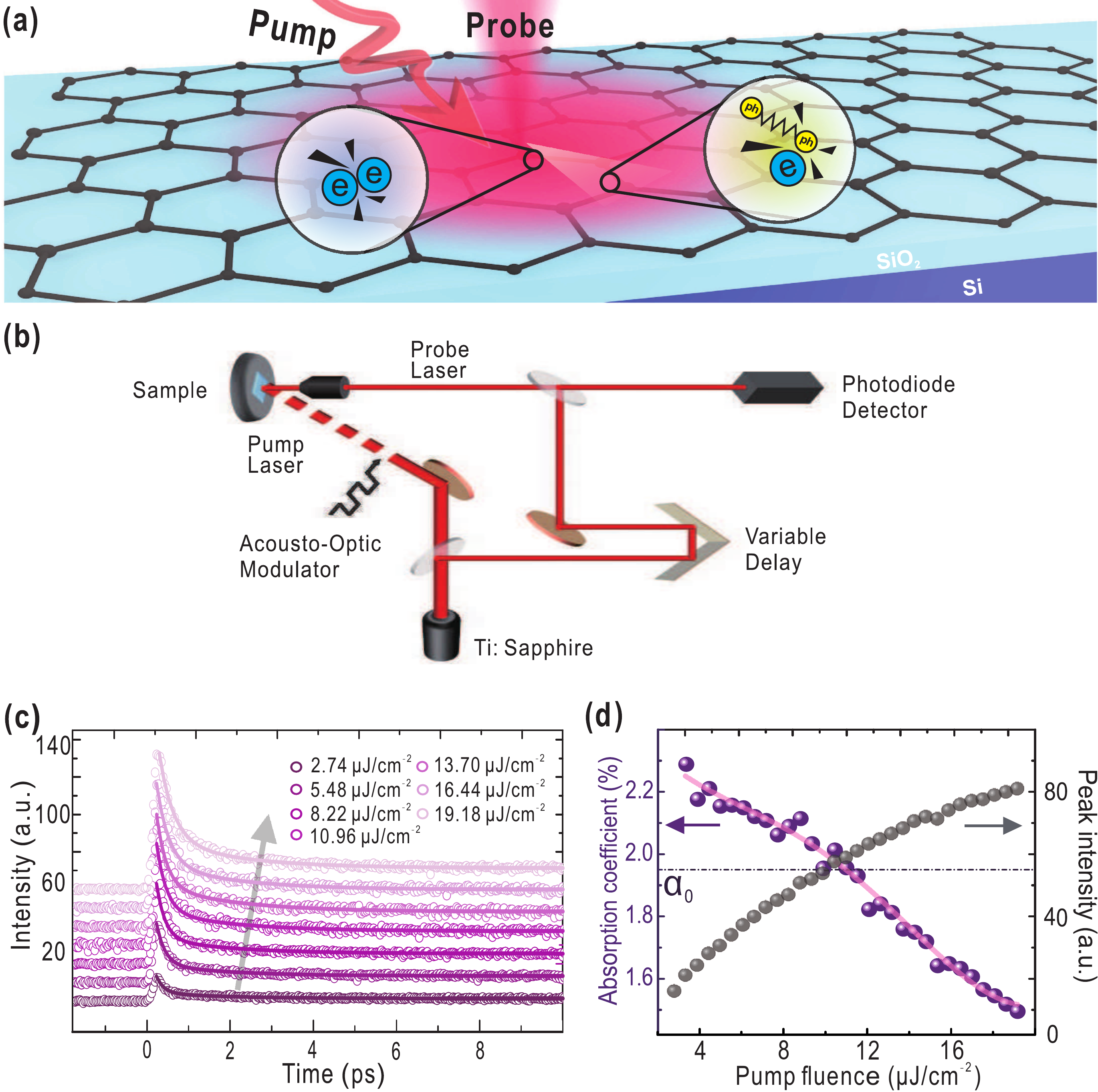}
\caption{(a) Schematic illustration of the monolayer graphene under ultrafast optical excitation. The optically-induced hot electrons cool down via \emph{e-e} scattering and \emph{e-ph} scattering. (b) Schematics of the setup for our femtosecond optical pump-probe measurement. (c) Delay-time dependence of the normalized transient differential reflection spectra measured for SG, while increasing the pump fluence from 2.74 $\mu J/cm^{-2}$ to 19.18 $\mu J/cm^{-2}$. The curves are shifted for clarity. (d) Right: the peak intensities of the measured signal in SG as a function of pump fluence (purple dots). Left: the deduced absorption coefficients of SG by experimental data (gray dots). Linear-absorption coefficient $\alpha_{0}$ for SG is marked by dashed line.}
\end{figure}

Graphene is a single atomic layer of $sp^{2}$ hybridized carbon with a honeycomb lattice, and is usually described in terms of massless Dirac fermions with linear dispersion near the Fermi level \cite{25,26,27,28,29}. The unique gapless band structure gives rise to an extremely short cooling time and ultra-broad nonlinear absorption ranging from visible to infrared region. Here, we start from a single layer graphene (SG) transferred on standard $Si/SiO_{2}$ substrate. As schematically illustrated in Fig.$\ $1(a), the optically-induced hot electrons cool down via scattering. The cooling down process usually includes carrier-carrier scattering and carrier-phonon scattering \cite{30,31}. To trace the carrier dynamics, femtosecond optical pump-probe measurements are employed (as shown in Fig.$\ $1(b)). Pump and probe pulses come from a Ti:sapphire mode-locked laser (Spectra-Physics Mai Tai HP) with pulse repetition rate as 80 MHz and center wavelength as 800 nm. Samples are illuminated by tightly focused pump pulses, and detected by a time delayed probe pulse with the same wavelength. Transient differential reflection ($\Delta R/R_{0}$) spectrum is measured as a function of delay time, which is defined as the arrival time of the probe pulse with respect to the pump pulse. All the measurements are carried out at room temperature.

Firstly, we explore the transient absorption of SG experimentally under different pump fluences. Figure 1(c) shows the measured transient differential reflection signal ($\Delta R/R_{0}$) of SG, the corresponding pump fluence increases from 2.74 $\mu J/cm^{-2}$ to 19.18 $\mu J/cm^{-2}$.  On each transient differential reflection spectrum, two distinct processes can be identified: one is the excitation of photo-induced hot electrons, which is characterized by the rapid increase of the signal intensity; the other is the cooling of hot electrons, which is characterized by the significant decay of intensity signal immediately after photoexcitation with a time scale of a few picoseconds. By increasing pump fluence, the signal intensity increases gradually. As shown in Fig.$\ $1(d), the extracted peak intensities of the signal possess a nonlinear relation with the pump fluence. Considering that the intensity of $\Delta R/R_{0}$ is proportional to the density of photo-induced electrons \cite{32}, we can directly obtain the evolution of the density profile of hot electrons over time. The absorption coefficient $\alpha \left( {I} \right)$ at different pump fluences is defined as the ratio between the induced hot electrons with certain pump energy. As plotted in Fig.$\ $1(d), the absorption coefficient are retrieved from the experimental data. The details are given in Sec.1 of the Supplemental Material. It follows that at different pump fluences, the experimental absorption coefficient $\alpha \left( {I} \right)$ is obviously different from the linear-absorption coefficient $\alpha_{0} \approx 1.95 \%$ \cite{33,34,35}, mainly due to the fact that the peak intensities of $\Delta R/R_{0}$ vary nonlinearly with the pump fluences. It is noteworthy that $\alpha \left( {I} \right)$ does not follow the two-level model \cite{4} based on Pauli blockade. Accroding to the two-level model, the absorption coefficient processes a maximum value ($\alpha_{0}$), and it always decays from $\alpha_{0}$ due to Pauli blocking \cite{36,37}. In our experiments, however, the absorption coefficient $\alpha \left( {I} \right)$ is much higher than $\alpha_{0}$ when the pump fluence is lower than a certain value, as illustrated in Fig.$\ $1(d). In other words, our data demonstrate that the Pauli blockade has been broken at low pump fluence in the optically-pumped graphene.

\begin{figure}[hbt]
\centering
\includegraphics[width=0.48\textwidth]{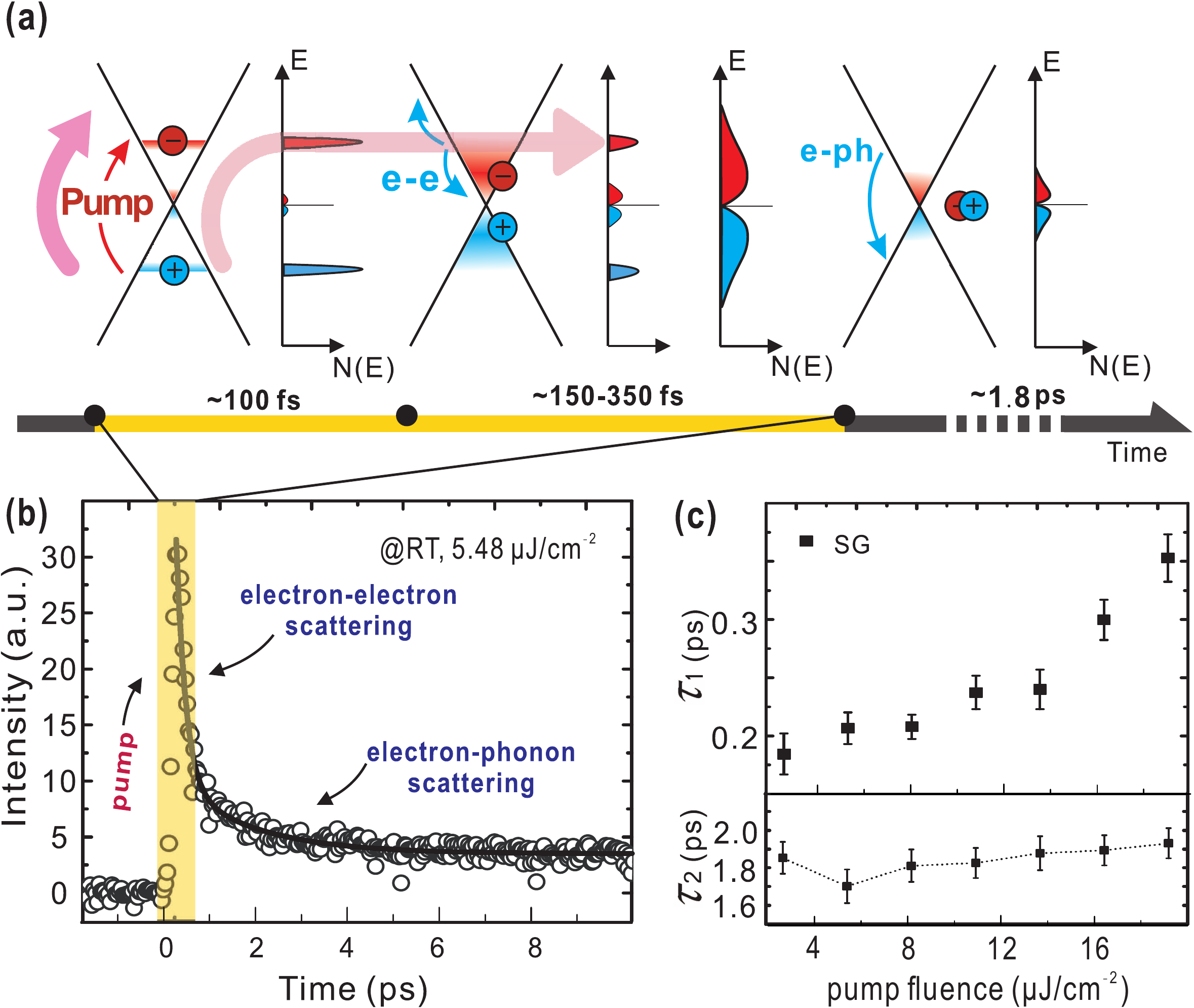}
\caption{ (a) Schematics of the evolution of photo-induced hot carriers after ultrafast optical excitation. Optical excitation process is indicated by the solid red arrow. Electrons are pumped from valence bands to the empty conduction bands with high energies. Then the hot electrons are cooling via \emph{e-e} and \emph{e-ph} scatterings, which are indicated by the solid blue arrows respectively. (b) The corresponding transient differential reflection spectrum measured, placing at the same time axis with (a). The black solid curve plotted is analytical fitted to the data using bi-exponentials function. (c) The \emph{e-e} scattering time ($\tau_{1}$) and \emph{e-ph} scattering time ($\tau_{2}$) of SG under various pump fluences respectively.}
\end{figure}

To understand the microscopic mechanism of breaking Pauli blockade in optically-pumped graphene, we explore the evolution processes based on the time-domain photo-induced hot electron dynamics. As schematically illustrated in Fig.$\ $2(a), electrons are pumped from valence bands to the empty conduction bands with high energies, generating photo-induced hot electrons. The measured transient differential reflection spectrum is shown in Fig.$\ $2(b). In the excitation stage, the transient differential reflection signal $ I_{0} \left( t \right) $ can be fitted by a single exponential function as \cite{38}
\begin{equation}\label{1}
 I_{0} \left( t \right) = A_{0}e^{t/\tau_{0}}+C,
\end{equation}
where $A_{0}$ is the initial signal amplitude, $\tau_{0}$ is the rise time, and $C$ is a constant. For all our samples in this work, the rise time $\tau_{0}$ is of the order of 100 fs, which is determined by the pulse duration of our laser source. After photoexcitation, the non-equilibrium carrier distribution broadens through rapid electron-electron (\emph{e-e}) scattering.  Consequently, those hot electrons cool down, yet still follow the Fermi-Dirac distribution with a temperature much higher than the lattice temperature. The electron-phonon (\emph{e-ph}) intraband scattering process contributes to further cooling of the hot electrons subsequently. Therefore in a real cooling process, the transient differential reflection signal $ I\left( t \right) $ can be fitted [shown as the solid curve in Fig.$\ $2(b)] by a bi-exponentially decaying function as \cite{39}
\begin{equation}\label{2}
 I \left( t \right) = A_{1}e^{-t/\tau_{1}}+A_{2}e^{-t/\tau_{2}}+C',
\end{equation}
where $ I\left( t \right) $ stands for the time-dependent signal in the cooling processes, $\tau_{1}$ and $\tau_{2}$ represent the cooling time of the \emph{e-e} and \emph{e-ph} scattering processes; $A_{1}$ and $A_{2}$ are the contribution weights of the two processes, respectively, and $C'$ is a constant. By fitting all curves in Fig.$\ $1(c) with Eq. (2), we obtain both $\tau_{1}$ and $\tau_{2}$ at different pump fluences. The details are given in Sec.2 of the Supplemental Material. As indicated in Fig.$\ $2(c), $\tau_{1}$ increases from around 180 fs to 350 fs when the pump fluence is increased from 2.74 $\mu J/cm^{-2}$ to 19.18 $\mu J/cm^{-2}$ (upper panel), which is of the same order of the time ($\tau_{0}$) for exciting an electron; whereas $\tau_{2}$ remains almost a constant around 1.8 ps (lower panel).  The proportional increment of $\tau_{1}$ with the pump fluence is due to the fact that the population of hot electrons increases at higher pump fluence, which prolongs the cooling of hot electrons.

The experimental data show that the electron excitation and the cooling down induced by \emph{e-e} interaction possess the same time scale [highlighted by the yellow region in Fig.$\ $2(b)] specifically at low pump fluence. The associated physical process can be understood as the following. When a femtosecond laser beam shines on a sample, the electrons are pumped to the empty conduction bands with high energies within $\sim$100 fs. The \emph{e-e} scattering cools down those hot electrons, which evacuates the conduction bands within the same time scale. Thereafter the electrons can be excited again. The electron excitation and hot-electron ultrafast cooling processes are alternatingly cycling. Hence excitation and cooling of hot electrons are strongly interplayed in hundred-femtosecond time scale. In this way, the empty conduction bands are ultrafast populated-depopulated repeatedly. Eventually Pauli blockade is dismissed, leading to an unconventionally enhanced optical absorption.

\begin{figure}[hbt]
\centering
\includegraphics[width=0.48\textwidth]{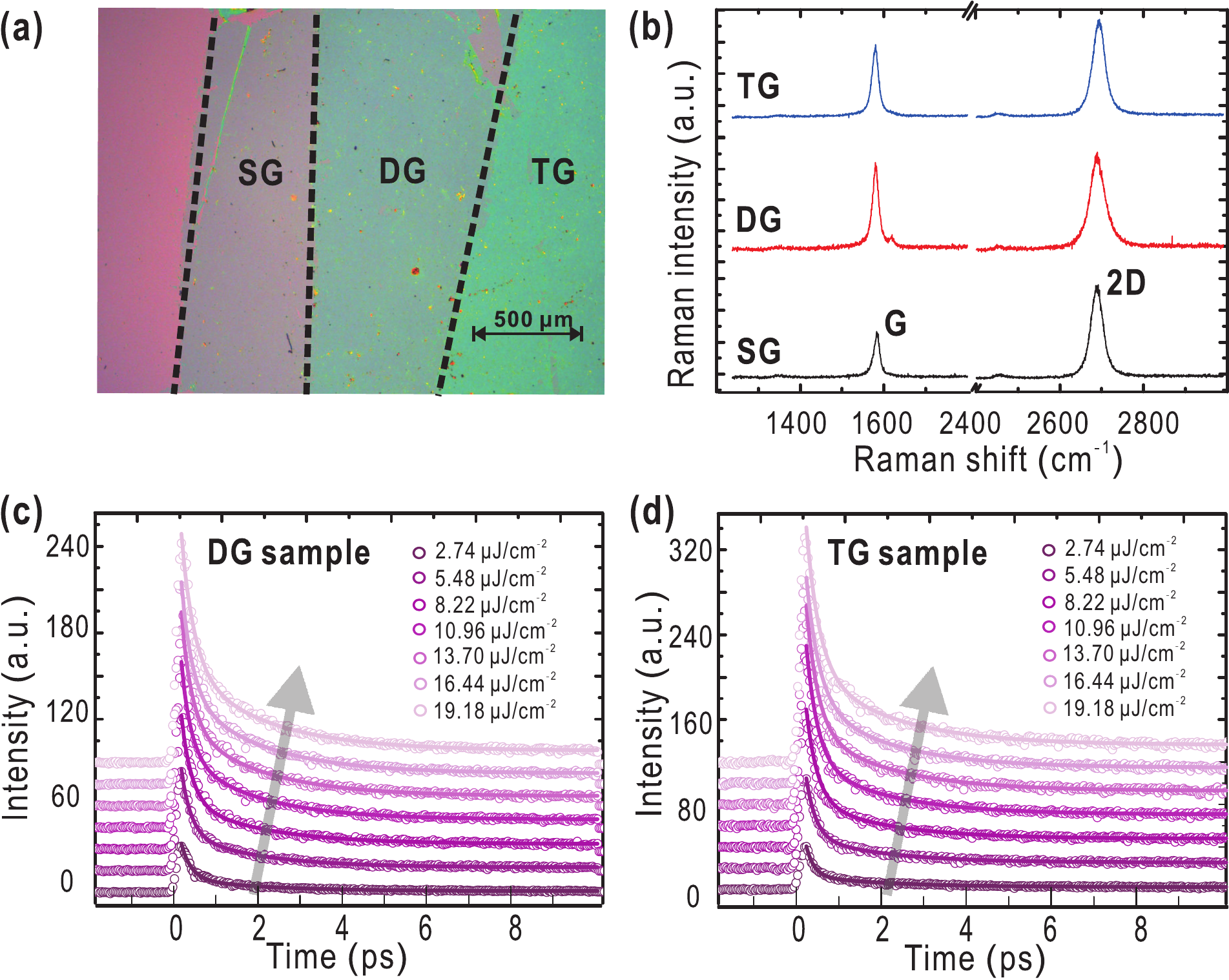}
\caption{(a) Optical image of SG, DG and TG transferred on $SiO_2/Si$ substrate. (b) Raman spectra for different stacking graphene layers, black, red and blue curves stand for SG, DG and TG, respectively. (c)-(d) The measured transient differential reflection signal verses delay time are plotted when increasing the pump fluence from 2.74 $\mu J/cm^{-2}$ to 19.18 $\mu J/cm^{-2}$. The corresponding solid curves are fitted to the experimental data for (c) DG sample and (d) TG sample, respectively.}
\end{figure}

\begin{figure*}[hbt]
\centering
\includegraphics[width=0.7\textwidth]{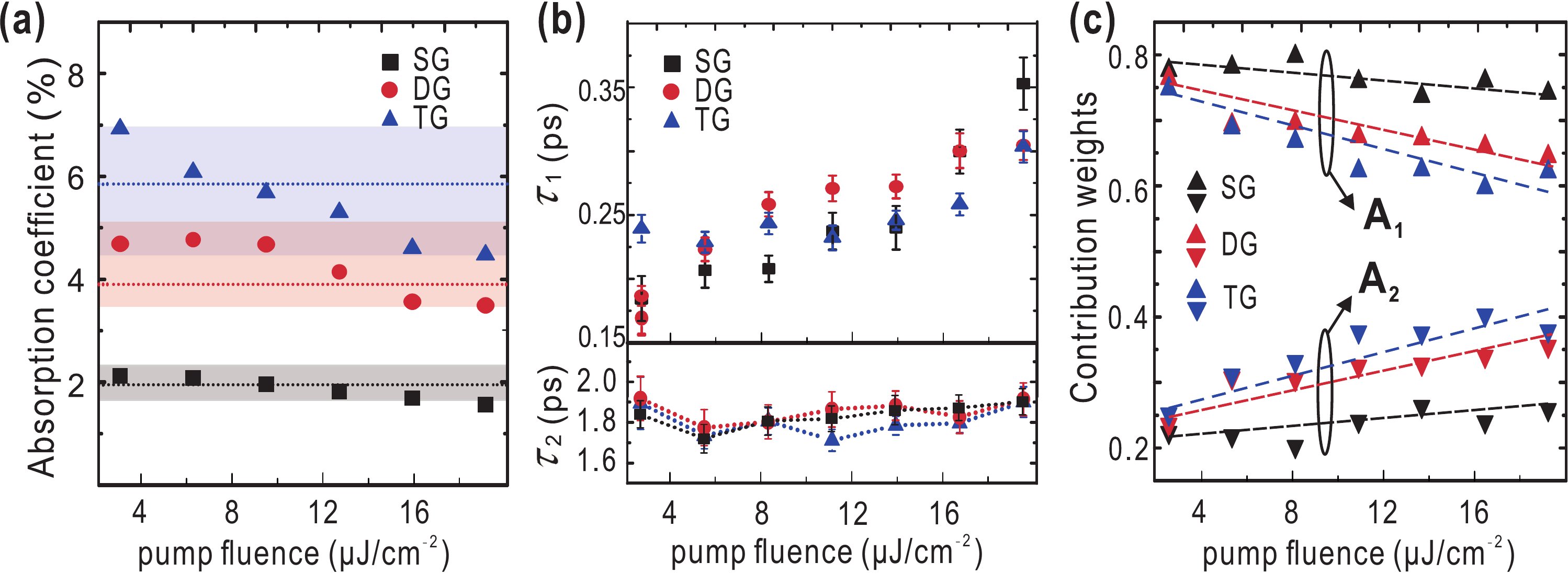}
\caption{ (a) Pump fluence dependence of the absorption coefficients. The linear absorption coefficients are marked by dashed lines with corresponding colors. (b) The \emph{e-e} scattering time ($\tau_{1}$) and the \emph{e-ph} scattering time ($\tau_{2}$) are plotted with varying pump fluences in the upper and lower panel individually. Gray, red and blue represent for SG, DG and TG respectively. (c) The contribution weights in the \emph{e-e} scattering process ($A_{1}$) and the \emph{e-ph} scattering process ($A_{2}$) are plotted at different pump fluences. Gray for SG, red for DG, and blue for TG.}
\end{figure*}

To strengthen the effect of breaking Pauli blockade, we increase the participated hot electrons by introducing double stacking graphene layers (DG) and triple stacking graphene layers (TG), respectively. Meanwhile, the number of hot electrons is nearly doubled (or tripled) in DG (or TG) compared with the scenario of SG under the same pump fluence. Figure 3(a) shows the optical microscopic images of SG, DG and TG. Experimentally we can easily distinguish the sample by color due to the fact that different layered sample has different optical absorption. Figure 3(b) shows the Raman spectra of graphene with different number of stacked layers with 514 nm laser excitation. The prominent G band at around 1580 $cm^{-1} $ and 2D band at around 2680 $cm^{-1}$ are clearly resolved in all samples, indicating that each layer still maintains the integrity of single layer graphene even after stacking. For monolayer graphene, the height of the 2D peak is almost twice as high as that of G peak. As the graphene layer increases, the ratio between the two peaks decreases, whereas the 2D peak remains symmetric and keeps the Lorentz profile \cite{40}.

We carry out femtosecond optical pump-probe measurements on DG and TG samples as well. The corresponding transient differential reflection spectra are shown in Figs.$\ $3(c) and $\ $3(d), respectively, which possess the similar features as that obtained from SG in Fig. 1(c). As the stacking layer increases, the signal intensity of the transient differential reflection increases since more hot electrons have participated in the ultrafast excitation and cooling processes. The absorption coefficients of DG and TG at different pump fluence are illustrated in Fig.$\ $4(a), which exhibit the same tendency as that of SG. It is noteworthy that in each scenarios, there always exists a value of critical pump fluence, below which the absorption coefficients become larger than the maximum value predicted by the two-level model, indicating that the Pauli blockade has been broken at the lower pump fluence. We have also determined the scattering time for DG and TG samples, respectively. As illustrated in Fig.$\ $4(b), the value of $\tau_{1}$ increases as the pump fluence is increased, whereas $\tau_{2}$ remains almost a constant. We can also figure out the contribution weights of the \emph{e-e} ($A_{1}$) and \emph{e-ph} ($A_{2}$) scattering processes. From Fig.$\ $4(c) we find $A_{1}$ is much larger than $A_{2}$ at low pump fluence, suggesting that the channel for electron cooling is dominated by \emph{e-e} scattering at low pump fluence. As the pump fluence increases, $A_{1}$ of the TG varies more rapidly, corresponding to a more evident change of the absorption coefficient $\alpha \left( {I} \right)$.

\begin{figure}[hbt]
\centering
\includegraphics[width=0.48\textwidth]{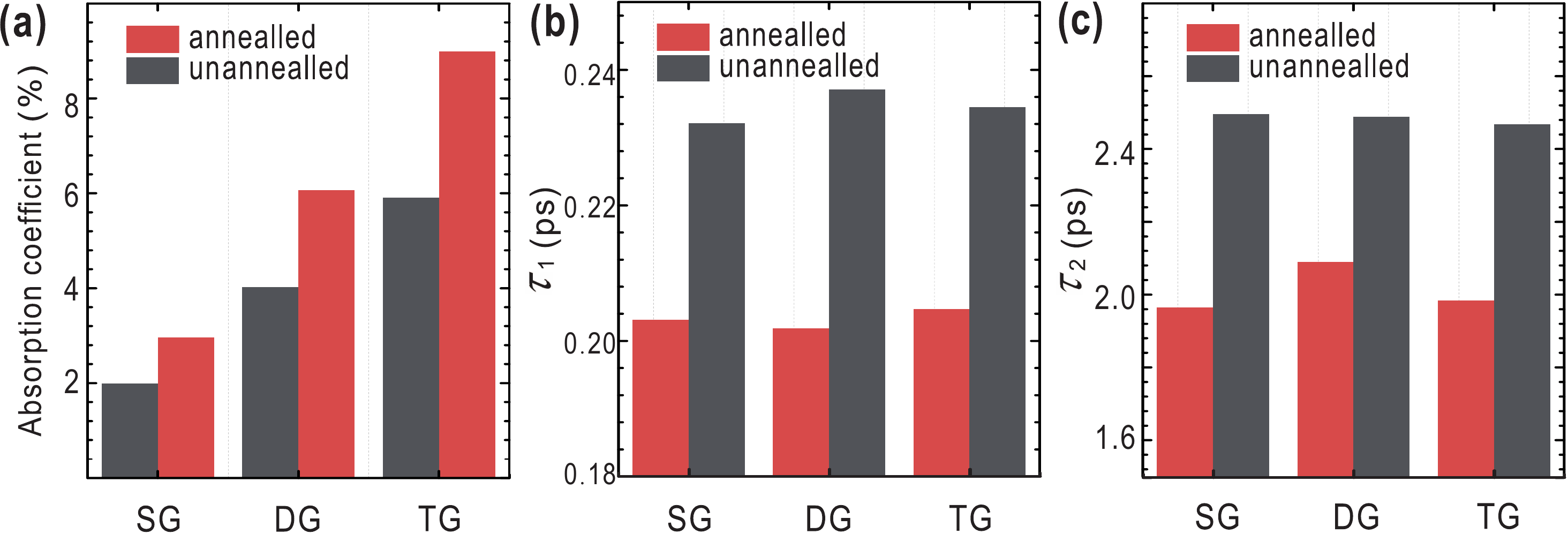}
\caption{ (a) The absorption coefficient of SG, DG, and TG samples under applying 5.48 $\mu J/cm^{-2}$ pump fluence, respectively, before or after annealing in an $Ar/H_{2}$ environment at 200${}^\circ C$ for 2.5 hours. (b) The \emph{e-e} scattering time ($\tau_{1}$) of SG, DG, and TG samples, respectively, before or after annealing. (c) The \emph{e-ph} scattering time ($\tau_{2}$) of SG, DG, and TG samples, respectively, before or after annealing.}
\end{figure}

An alternative approach to enhance the ultrafast cooling of hot electrons is to apply annealing pretreatment to samples. Recently, thermal annealing has been investigated as a common practice to eliminate contamination and restore clean surfaces of two-dimensional materials, and the annealed samples attach more tightly on the substrate \cite{41}. Here we compare the optical absorption and electron cooling behaviors on the samples with and without annealing. Two types of graphene samples have been prepared: one is pristinely transferred to substrate without any subsequent thermal treatment; whereas the other is annealed in an $Ar/H_{2}$ environment at 200${}^\circ C$ for 2.5 hours. By applying 5.48 $\mu J/cm^{-2}$ pump fluence, the absorption coefficient of the annealed sample (\emph{i.e.}, SG, DG, and TG, respectively) shows evidently a rather larger value comparing to the pristine sample, as illustrated in Fig.$\ $5(a). Moreover, \emph{e-e} scattering time has been shortened from 230 fs to 200 fs (Fig.$\ $5(b)), and \emph{e-ph} scattering time has been shortened from around 2.4 ps to 2.0 ps as well (to see Fig.$\ $5(c)). Physically, annealing substantially reduces the charged impurities and elininates structural disorders, and therefore enhances the carrier mobility in graphene \cite{42,43}. We can find that via the annealing, both \emph{e-e} scattering and \emph{e-ph} scattering are effectively enhanced. Yet \emph{e-ph} scattering (lasting for a few picoseconds) does not directly contribute to the breaking of Pauli blockade according to our experimental data at different substrate temperatures (the details are given in Sec.3 of the Supplemental Material), which do agree with the aforementioned microscopic picture of breaking Pauli blockade in the optically-pumped graphene. By annealing pretreatment, \emph{e-e} scattering time has been shortened to femtosecond time level, which does help to evacuate the conduction bands quickly to maintain the ultrafast populating-depopulating process. The consequence of this process is that the absorption coefficient goes beyond the limit predicted by the two-level model at low pump fluence. All these analysis indicate that we indeed enhance the ultrafast cooling of hot electrons by sample annealing pretreatment. The cooperative populating-depopulating of the conduction band leads to the breaking of Pauli blockade.

In conclusion, we demonstrate here the observation that the ultrafast cooling of hot electrons can break Pauli blockade in optically pumped graphene. The ultrafast pump-probe experiments are carried out to monitor the dynamics of photo-induced hot electrons down to femtosecond. The time scale for electron excitation is proven to be of the same order of electron cooling ($\sim$100 fs) through \emph{e-e} scattering. Consequently, ultrafast electron excitation interplays with hot electron cooling, leading to repeated population-depopulation of the empty conduction bands. This effect eventually dismisses the Pauli blockade, contributing to a significantly increased optical absorption. Despite that this unconventional effect is verified in graphene, we foresee that it is not restricted to graphene-based system only. It can exist in other materials with simple energy level scheme. We anticipate that this microscopic view provides valuable perspectives for exploring the next generation of high-speed ultrathin photonic devices.

\end{document}